\documentclass[fleqn,usenatbib]{mnras}

\usepackage[T1]{fontenc}
\usepackage{ae,aecompl}

\usepackage{graphicx}   
\usepackage{amsmath}    
\usepackage{amssymb}    

\newcommand{\km}{\rm\thinspace km}

\newcommand{\erg}{\rm\thinspace erg}
\newcommand{\s}{\rm\thinspace s}
\newcommand{\Mpc}{\rm\thinspace Mpc}

\newcommand{\kmps}{\hbox{$\km\s^{-1}$}}
\newcommand{\ergps}{\hbox{$\erg\s^{-1}\,$}}

\newcommand{\chisq}{\hbox{$\chi^2$}}

\newcommand{\kmpspMpc}{\hbox{$\kmps\Mpc^{-1}$}}
\newcommand{\civ}{\ion{C}{iv}}
\newcommand{\mgii}{\ion{Mg}{ii}}
\newcommand{\kelvin}{\rm\thinspace K}
\newcommand{\K}{\rm\thinspace K}

\title[Hot Dust and Quasar Outflows]{Exploring the link between C\,{\Large \textbf{IV}} outflow kinematics and sublimation-temperature dust in quasars}

\author[M. J. Temple et al.]{Matthew J. Temple,$^{1,2}$\thanks{E-mail: matthew.temple@mail.udp.cl}
Manda Banerji,$^{3,1,4}$
Paul C. Hewett,$^{1}$
Amy L. Rankine,$^{1}$
\newauthor
and Gordon T. Richards$^{5}$
\\
$^{1}$Institute of Astronomy, University of Cambridge, Madingley Road, Cambridge CB3 0HA, UK\\
$^{2}$N\'ucleo de Astronom\'ia, Universidad Diego Portales, Av. Ej\'ercito Libertador 441, Santiago, Chile\\
$^{3}$School of Physics \& Astronomy, University of Southampton, Southampton, SO17 1BJ, UK\\
$^{4}$Kavli Institute for Cosmology, University of Cambridge, Madingley Road, Cambridge CB3 0HA, UK\\
$^{5}$Department of Physics, Drexel University, 32 S. 32nd Street, Philadelphia, PA 19104, USA
}

\date{Accepted XXX. Received YYY; in original form ZZZ}

\pubyear{2020}

\begin{document}
\label{firstpage}
\pagerange{\pageref{firstpage}--\pageref{lastpage}}
\maketitle

\begin{abstract}
Using data from SDSS, UKIDSS and {\it WISE}, we investigate the properties of the high-frequency cutoff to the infrared emission in $\simeq$5000 carefully selected luminous ($L_{\rm bol}\sim10^{47}$) type 1 quasars.
The strength of $\simeq$2\,\micron\ emission,  corresponding to  emission from the hottest ($T>1200\K$) dust in the sublimation zone surrounding the central continuum source,
is observed to correlate with the blueshift of the \ion{C}{IV} $\lambda$1550 emission line.
We therefore find that objects with stronger signatures of nuclear outflows tend to have a larger covering fraction of sublimation-temperature dust.
When controlling for the observed outflow strength, the hot dust covering fraction does not
vary significantly across our sample as a function of luminosity, 
black hole mass or Eddington fraction.
The correlation between the hot dust and the \ion{C}{IV} line blueshifts, together with the lack of correlation between the hot dust and other parameters, 
therefore provides evidence of a link between the properties of the broad emission line region and the infrared-emitting dusty regions in quasars.

\end{abstract}




\section{Introduction}

It has been known for decades that quasars, and other AGN, show an excess of emission in the rest frame near-infrared at $\sim$1-3\micron. This emission is  attributed to dust that has been heated to its sublimation temperature by the central continuum source \citep{Rees69, Rieke78, Barvainis87}.
Interferometric observations \citep{2009A&A...507L..57K, 2011A&A...527A.121K, 2020A&A...635A..92G} confirm that the emission from the shortest-wavelength components of the near-infrared excess is located close in to the central continuum source, while reverberation mapping studies have shown that the hot dust is located just outside the broad emission line region (BLR), perhaps even setting the outer limit of that region
\citep{2006ApJ...639...46S, 2012MNRAS.426.3086G, 2019MNRAS.489.1572L, 2019ApJ...886..150M}.

The sublimation temperature of astrophysical dust in AGN is largely set by its composition, with little change expected for physically reasonable variations in the shape of the ionizing spectral energy distribution (SED) or the size of the dust grains: pure silicate dust sublimates at temperatures around 1300-1500\kelvin, and pure graphite dust at 1800-2000\kelvin.
This narrow temperature range allows a prediction to be made for the sublimation radius, which is larger than light-travel distance corresponding to the hot dust time lag found from reverberation mapping.
These shorter-than-expected time lags suggest that the hot dust is located above the plane of the  accretion disc, leading to a geometric foreshortening effect
(e.g. \citealt{2012MNRAS.426.3086G}, fig. 1; \citealt{2020AJ....159..259F}, fig. 14).

Over the past three decades, several authors have proposed models which use disc-driven winds to account for the dusty toroidal structures which obscure the inner regions of AGN from certain viewing angles \citep[e.g.][]{1994ApJ...434..446K, 2019ApJ...884..171H}. More recently, models have been proposed which link these winds, or more specifically, associate the failed parts of these winds, with the broad emission line region \citep{2011A&A...525L...8C, 2017ApJ...846..154C, 2018MNRAS.474.1970B}. These models predict that the properties of the broad line region and the dust emitting regions should be related, however, observational studies to probe such links are somewhat small in number.
One such study is that of  \citet{Wang13}, who found that the rest-frame near-infrared colours in a sample of $\simeq$4700 quasars from the fifth data release of the Sloan Digital Sky Survey (SDSS) were correlated with the properties of the \civ\  $\lambda$1550 emission line. 
More specifically, they found that objects with broader and bluer \civ\ emission lines tend to show brighter emission at $\approx$3\,\micron\ in the 2012 All-Sky Data Release from \textit{WISE},
and that the strength of this correlation increases with increasing Eddington ratio.

Since 2013, the reactivated NEOWISE project has continued to scan the sky, leading to a large increase in the total amount of data gathered in the 3.4 (\textit{W1}) and 4.6\,\micron\ (\textit{W2}) bands. The unWISE coadds presented by \citet{2019PASP..131l4504M} make use of ten full-sky mappings in \textit{W1} and \textit{W2}, compared to the single mapping used to construct the 2012  \textit{WISE} All-Sky Data Release, with a corresponding decrease in the noise associated with each photometric measurement.
At the same time, the SDSS has continued to identify new objects: the fourteenth data release includes some 526\,356 quasars, compared to the  77\,429 from the fifth data release. This increase in sample size allows the hot dust properties to be quantified in a much larger number of quasars, and for the dependence of these parameters on the emission line properties {\em and} the Eddington fraction to be explored simultaneously. Recent work by \citet{Rankine20} has developed an improved scheme for masking absorption features within the SDSS spectra, leading to an improved determination of the emission line properties in each object. In particular, the reported blueshift of the \civ\ emission line is now a more robust measurement of the strength of emission from outflowing  gas.

The primary aim of this investigation is to quantify the properties of the high-frequency cutoff to the dust emission in luminous quasars, corresponding to emission from the hottest temperature dust, and to compare to the outflow signatures seen in the high ionization broad emission lines.
The rest frame near-, mid-, and far-infrared emission in quasars is due to several dust components at multiple temperatures: within the type-1 quasar population there exists a diversity of emission at $\lambda>3\,\micron$, with the cooler dust emission (at longer wavelengths) not necessarily correlating with the hottest dust emission. These cooler components can be  due to emission from dust at larger radii where the equilibrium temperature of the AGN-illuminated material is lower, and or due to dust in the host galaxy which has been heated by young stars.
For this reason we restrict this investigation to  emission at rest-frame wavelengths shorter than 3\,\micron, where the emitting dust has temperatures $T\gtrapprox1200\K$ and  must therefore be located near the inner regions of the AGN and  heated by the ultraviolet (UV)--optical continuum from the central engine.

The structure of this paper is as follows. In Section \ref{sec:data} we describe the 
{data from SDSS, UKIDSS and {\it WISE} used in this work.}
From these surveys we construct a sample of 
$\approx$12\,000  quasars at redshift $z\approx2$ with photometric data covering the rest frame 0.1-3\,\micron\ range.
Our method for quantifying the hot dust properties of this sample is described in Section~\ref{sec:method}. We compare the hot dust properties to the UV emission line properties in Section~\ref{sec:results} and discuss our results in Section \ref{sec:discuss}. 
We assume a flat $\Lambda$CDM cosmology with $\Omega_m=0.27$, $\Omega_{\Lambda}=0.73$, and $\textrm{H}_0=71 \kmpspMpc$. All emission lines are identified with their wavelengths in vacuum in units of \AA ngstr\"{o}ms. 
The rest frame monochromatic quasar luminosity at 3000\,\AA\ 
(hereafter $L_{3000}$) is used as a measure of the strength of the UV continuum.
Following \citet{Shen11}, we use the constant bolometric correction $L_{\rm bol} = 5.15 \times L_{3000}$ to estimate Eddington ratios, derived from the composite SED in \citet{Richards06}.

\section{Data}
\label{sec:data}

All the quasars used in this work are drawn from the Sloan Digital Sky Survey (SDSS) fourteenth data release (DR14) quasar catalogue, the selection of which is summarised in \citet{Paris18}. 
Some of these quasars were selected in part based on their mid-infrared {\it WISE} fluxes, and we show in Appendix~\ref{sec:select} that our results are not biased by the inclusion of these objects.

The spectroscopic and photometric data from SDSS together with photometric data from infrared sky surveys employed are described below.

\subsection{Photometric data}
\label{sec:photometry}

\begin{table}
 \centering
\caption{\small
The number of quasars  remaining at each stage of  the cross-matching of catalogues described in Section~\ref{sec:photometry}, and   construction of the three samples used in Sections~\ref{sec:tbb}, \ref{sec:primary} and Appendix~\ref{sec:highz}.
}
\begin{tabular}{l| l | r}
\hline
 &  & No. of quasars\\
\hline
Section~\ref{sec:photometry} & SDSS DR14 quasar catalogue & 526\,356 \\
 & matched to unWISE & 452\,315 \\
 & matched to UKIDSS-LAS & 105\,017\\
\hline
Section~\ref{sec:tbb}  & $1.2<z<1.5$ & 6\,910 \\
 & $Y<18.1$ and confident point sources & 1\,248  \\
 & detected in \textit{grizYJKW12} & 1\,213  \\
\hline
Section~\ref{sec:primary}  & $1.2<z<2.0$ & 18\,303 \\
 & $Y<19.3$ & 13\,327  \\
 & detected in \textit{grizYJKW12} & 12\,294  \\
 & \civ\ line information & 5\,022  \\
\hline
Appendix~\ref{sec:highz}  & $2.75<z<3.25$ and AllWISE & 4\,768 \\
 & $Y<18.3$ & 837  \\
 & detected in \textit{rizYJHKW123} & 701  \\
\hline
\end{tabular}
\label{tab:data}
\end{table}

We use {\it griz} point-spread-function magnitudes from the SDSS quasar catalogues \citep{Paris18, Schneider10}. Above $z\approx1.5$, the {\it u} band flux is affected by absorption from neutral hydrogen in the intergalactic medium (IGM), which dramatically reduces the amount of flux at wavelengths shorter than that of Ly\,{\rm$\alpha$}, and we exclude the {\it u} band from this work. The SDSS DR14  quasar catalogue \citep{Paris18} has spectra for 526\,356 quasars in the redshift range $0<z<5$.

{We start by cross-matching the} \citet{Paris18} catalogue to mid-infrared {\it WISE} data \citep{Wright10}; specifically the unWISE catalogue presented by \citet{Schlafly19}, who use the  \citet{2019PASP..131l4504M} coadds. 
unWISE has $\approx$0.7 magnitudes deeper coverage in {\it W1} and {\it W2} compared to AllWISE, corresponding to additional data from the reactivation of NEOWISE. 
We match the SDSS DR14 quasar catalogue to unWISE using a 3.0\,arcsec matching radius and  keep only sources with a unique match within that radius. 
unWISE models the {\it W1} and {\it W2} pixels separately, and to reduce the number of contaminants we keep only sources where there is a detection in both {\it W1} and {\it W2}. 
We find 452\,315 unWISE sources match to quasars from SDSS DR14.
452\,199 of these matches have signal-to-noise ratio (S/N)  greater than 5 in both {\it W1} and {\it W2}.
In Appendix~\ref{sec:WISE} we show that our results are unchanged when instead using AllWISE as the source of {\it WISE} data.

We then cross-match the SDSS DR14 quasar catalogue to the tenth data release of the UKIDSS Large Area Survey \citep[UKIDSS-LAS DR10;][]{Lawrence07}.
We use  `apermag3' values, which are the default point-source  2.0\,arcsec diameter aperture corrected magnitudes.
The UKIDSS-LAS covers approximately three tenths of the SDSS footprint in the near-infrared {\it YJHK} bands. We match 116\,370 quasars from SDSS DR14 to unique sources within 1.0\,arcsec in the UKIDSS-LAS DR10 catalogue. The number of detections is slightly larger than given by \citet{Paris18}, who use a force-photometry method, and we believe the improvement is due to our use of more recent data from UKIDSS.
Of the 452\,315 SDSS quasars which we match to unWISE, 105\,017 are also matched to the UKIDSS-LAS.
Restricting our sample to the UKIDSS footprint reduces the number of objects in our sample by some seven-tenths, however the inclusion of the K band in particular allows us to  better constrain the flux emitted from each object at wavelengths just short of 1\,\micron\ in their rest frame, and hence identify how much of the flux at $\lambda>1$\,\micron\ is due to the tail of the quasar UV--optical continuum, and therefore not due to hot dust.

We quote all infrared magnitudes on the Vega system, which is the native magnitude system for UKIDSS and \textit{WISE}. 
Optical magnitudes are given on the SDSS system.

\subsubsection{Galactic extinction} 
\label{sec:galext}

We correct the {\it grizYJHK} photometric bands for Galactic extinction using the improved 
 dust maps of \citet{Schlafly10} and \citet{Schlafly11}. 
Such extinctions are typically of the order $E(B-V)<0.1$, 
as most quasars in the SDSS footprint lie outside the Galactic plane.

Commonly quoted passband attenuations are correct only for sources with SEDs similar to that of a elliptical galaxy at low-redshift,
and we derive our own passband attenuations using a $z=2.0$ quasar source SED,
assuming $R_V=3.1$ and the Galactic extinction curve of \citet{Fitzpatrick09}.
A type-1 quasar SED is bluer than a typical stellar or galaxy SED, leading to subtle differences in the conversion from $E(B-V)$ to the attenuation in each observed passband.
The conversions we use are given in Table~\ref{tab:galext}.  
Using source quasar SEDs in the range $1.5\le z \le 2.5$ alters these values by no more than $1.5$\,per cent.
The attenuation due to dust in the {\it WISE} bands is found to be negligible for the extinctions we consider and no correction for Galactic extinction is applied to these data.

\begin{table}
 \centering
\caption{\small Passband attenuations $A_\lambda/E(B-V)$ for \textit{grizYJHK}-filters adopting the $R_V=3.1$ Galactic reddening law from \citet{Fitzpatrick09} and a $z=2.0$ quasar SED.}
\begin{tabular}{l |c }
\hline
Filter & $A_\lambda/E(B-V)$ \\
\hline
SDSS \textit{g} & 3.80\\
SDSS \textit{r} & 2.58\\
SDSS \textit{i} & 1.92\\
SDSS \textit{z} & 1.42\\
UKIDSS \textit{Y} & 1.12\\
UKIDSS \textit{J} & 0.80\\
UKIDSS \textit{H} & 0.52\\
UKIDSS \textit{K} & 0.33\\
\hline
\end{tabular}
\label{tab:galext}
\end{table}

\subsection{Spectroscopic data}

For objects included in the \citet{Schneider10} DR7 quasar catalogue, emission-line measurements are made using spectra \citep{2010arXiv1010.2500W} processed using the improved sky subtraction of \citet{2005MNRAS.358.1083W}.
For quasars identified after SDSS DR7, emission-line measurements are derived from the spectra presented in \citet{Paris18}. 

\subsubsection{\civ\ line measurements}

\civ\ $\lambda1550$ is a high ionization line (64 eV), where the 
line profile is often skewed with an excess of emission in the blue wing. This asymmetry is believed to be due to outflowing material moving along the line of sight towards the observer, possibly in some wind driven off the accretion disc. Blue excesses can skew the location of the median line flux by up to $\approx$5000\kmps\ relative to the systemic redshift \citep{Richards11}.

However, the \civ\ emission line is often contaminated by absorption features and so we calculate emission-line properties using the spectral reconstructions presented by \citet{Rankine20}.
Briefly, these reconstructions use the known correlation between the morphology of the `\ion{C}{iii}]'\,$\lambda1909$-emission complex and the blueshift of \civ\ to place priors on the weights of linear spectral components derived from a mean field independent component analysis of quasars which do not have any absorption features. 
These components are fit to the unabsorbed pixels in each quasar spectrum, where absorbed pixels are identified and masked in an iterative fashion until the fit converges and no unmasked pixel is more than $2\sigma$ below the reconstruction. The reconstructions essentially provide a higher S/N version of each spectrum, with both broad and narrow absorption features masked to allow a more robust measurement of the emission line properties.

We use the \civ\ line properties measured by \citet{Rankine20}, who follow the prescription described by \citet{Coatman16, Coatman17}. A power law in flux is defined using the wavelength windows 1445-1465\,\AA\ and 1700-1705\,\AA, and subtracted from the spectrum over 1500-1600\,\AA. The line `blueshift' is then defined as the Doppler shift of the line centroid relative to the rest frame wavelength of the emission line at the systemic redshift:
\begin{equation}
    {\rm \civ\ blueshift} \equiv c\times (\lambda_{\rm rest}-\lambda_{\rm median})/\lambda_{\rm rest}
\end{equation}
where $\lambda_{\rm rest}=1549.48$\,\AA\ is the rest frame wavelength of the \civ\ doublet\footnote{Assuming equal contributions in emission from both components.} and $\lambda_{\rm median}$ is the wavelength bisecting the total continuum-subtracted line flux.
The \civ\ blueshift is effectively a measure of the balance of `wind'  to `core' emission in a model where  \civ\  is emitted from clouds of gas which are either outflowing in a wind or located at the systemic redshift.

We exclude all objects which show broad low-ionization absorption features (LoBALs), leaving 141\,488 quasars with \civ\ information in the redshift range $1.56<z<3.50$.

\subsubsection{Systemic redshifts}
\label{sec:redshift}
We calculate systemic redshifts using a mean-field independent component analysis scheme 
(Allen \& Hewett in prep.) with the redshift as a free parameter. 
This scheme excludes the \civ\ emission line from the analysis to avoid biasing the systemic redshifts with information from skewed emission lines.
These redshifts are the same as those presented in \citet{Rankine20} for those quasars which have \civ\ line measurements.

\subsubsection{Black hole masses}
 The asymmetry in \civ\ emission shows that at least some of the emitting gas is not virialised and deriving unbiased black hole masses from \civ\ is not straightforward \citep[][]{Coatman17}. We instead use the \mgii\ $\lambda2800$ emission line to estimate black hole masses.

Black hole masses are estimated using the Full Width at Half Maximum (FWHM) of the \mgii\ line and the single-epoch virial estimator described by \citet{Vestergaard09}, having subtracted \ion{Fe}{ii} emission using the template of \citet{Vestergaard01}:
\begin{equation}
    M_{\rm BH} = 10^{6.86}
    \left(\frac{{\rm FWHM(\mgii)}}{1000\kmps}\right)^2
    \left(\frac{L_{3000}}{10^{44}\ergps}\right)^{0.5}M_\odot
\end{equation}
Eddington ratios are calculated assuming $L_{\rm bol} = 5.15 \times L_{3000}$. In Section~\ref{sec:BC}
we discuss the assumption of a constant bolometric correction and the possibility that it should perhaps be changing as a function of \civ\ blueshift.

\section{Method}
\label{sec:method}

\begin{figure*}
    \includegraphics[width=2\columnwidth]{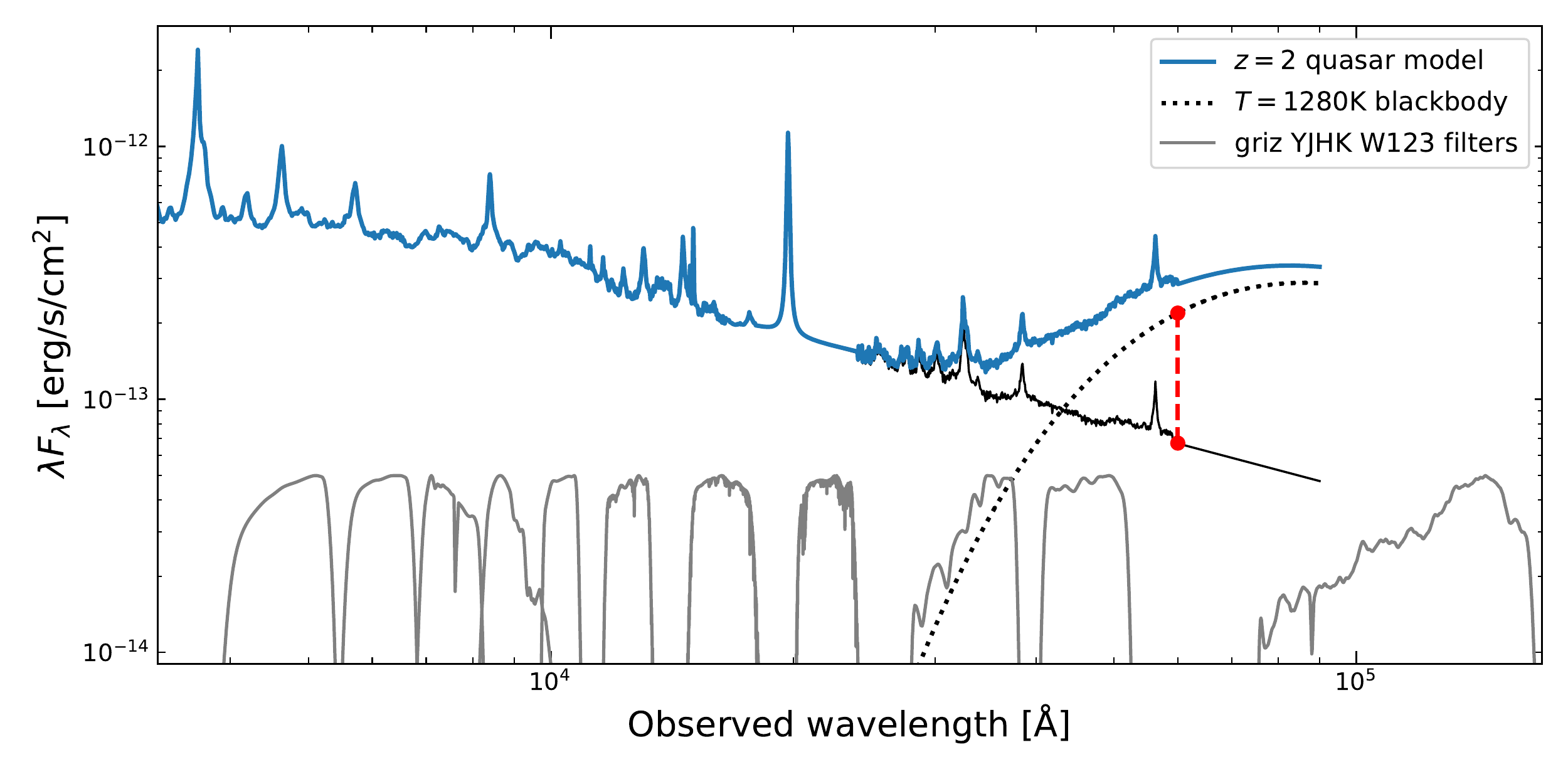}
    \caption{An example of our quasar model with $z=2.0$, $E(B-V)=0.0$, $L_{3000}=10^{46}\ergps$, and $L_{\small 2\,\micron, \textrm{HD}}/ L_{\small 2\,\micron, \textrm{QSO}}=4$. We quantify the near-infrared excess in the best-fitting model for each object in our sample using the ratio of the luminosities at ${2\,\micron}$ in the rest frame (dashed red line) of the hot dust blackbody (dotted black) and quasar power-law continuum (solid black).
    The normalised SDSS, UKIDSS and \textit{WISE} filter response curves are shown in grey.
    {When fitting the model to the observed photometric colours of individual objects, the only free parameters control the amount of extinction and the hot dust blackbody parameters.}
    }
    \label{fig:model}
\end{figure*}

\subsection{Parametric model}
\label{sec:model}

We use a parametric SED model to describe  quasar emission in the 1200\,\AA\ to 3\micron\ rest-frame wavelength range.
This model will be presented in Temple et al. (2021 in prep.),
where we will set out to reproduce the colours of SDSS DR16 quasars across a wide range of luminosities and redshifts.
For the purposes of this work, we have adopted model parameters which best describe the median \textit{ugrizYJHKW12} colours of a subset of DR14 quasars, where the subset was chosen to match the luminosities of the objects considered in this work.
The model is built using a broken power law with slopes $f_\lambda \propto \lambda^{-1.52}$ and  $f_\lambda \propto \lambda^{-1.84}$ for $\lambda < 2820$ and $\lambda > 2820$\,\AA\ respectively.
Emission lines are included using templates derived from the composite spectra presented in
\citet{Francis91} and \citet{Glikman2006},
with the power-law slopes and the strength of the emission lines chosen to match the median DR14 colours.
The near-infrared component is modelled by a blackbody where the temperature and normalisation are both free to vary.The quasar SED model is very similar to that used by \citet{Maddox08, Maddox12}, and an example SED is shown in Fig.~\ref{fig:model}.

The total model can be reddened using a quasar extinction curve with the $E(B-V)$ as a free parameter.
The extinction curve is the same as used in \citet{Temple19} and \citet{Wethers18}.
However, we do not interpret the modest range of derived $E(B-V)$ values as primarily resulting from dust reddening; rather it introduces flexibility to allow the model to account for slight variations in the quasar continuum slope. We thus allow the $E(B-V)$ to take negative values to account for objects which have bluer continua than average.

For this work, the model can be thought of as a function which takes four parameters (redshift, $E(B-V)$, blackbody temperature and blackbody strength) as inputs and returns predicted colours for any given combination of photometric filters.
Every object considered in this work is a spectroscopically confirmed quasar and we fix the redshift for each object using the spectroscopic redshifts described in Section~\ref{sec:redshift}.

We note that the $\simeq$1\,\micron\ inflexion point in the quasar SED is sensitive to photospheric stellar emission from the host galaxy. In particular, objects with an excess of host galaxy emission which is otherwise unaccounted for will require a progressively hotter blackbody to adequately account for the shape of the observed SED at wavelengths $\ge$1\,\micron.
However, due to the limited number of photometric data points which we have available to use in the rest-frame near-infrared, the strength of the host galaxy emission and the temperature of the hot dust component are somewhat degenerate. We therefore chose not to include a host galaxy component in the model SED, but instead mitigate against this uncertainty by restricting our sample to objects which are classed as point sources in all UKIDSS bands, and impose additional flux limits to exclude fainter objects which are more likely to have a significant contribution from starlight.
We show in Appendix~\ref{sec:host} that our results remain unchanged when including a component in our model to account for starlight from the host galaxy.

\subsection{Fitting routine}

For a given redshift, $E(B-V)$, blackbody temperature and blackbody strength, we generate a model SED.
We multiply that SED by the \textit{grizYJHKW12} filter response curves to obtain synthetic photometric colours.
These are fit to the observed colours using a Nelder-Mead algorithm.
A floor of 0.05 mag is imposed on the photometric uncertainties, corresponding to a minimum flux uncertainty of 5\,per cent.

A simpler approach would be to treat each photometric passband as a monochromatic data point and fit a combination of power laws to those data points \citep[e.g.][]{Wang13}. 
However,  as the redshift of the source increases, the slope of the observed SED at $\simeq$2-4\,\micron\ will change as the inflexion point moves to longer observed-frame wavelengths. 
The change in SED slope leads to a change in the effective wavelength of each passband $\lambda_{\rm eff}$, and thus an additional uncertainty in the parameters of the best fitting model.
The benefit of our approach, therefore, is that it avoids introducing additional redshift-dependant scatter in the measurement of the strength of the hot dust emission.

The strongest emission line in the wavelength range of the type-1 quasars we consider is H${\rm \alpha}$, variation of which can change the total flux in an observed passband by up to $\simeq$0.2 mag. We therefore exclude any band which includes H${\rm \alpha}$ line at the redshift of a quasar. In practice this means that, for objects at redshift $1.2<z<2.0$, we exclude the $J-H$ and $H-K$ colours and instead include the $J-K$ colour when fitting our model to the data. All other optical and near-infrared emission lines have much smaller equivalent widths and the flux in all the other filters is dominated by continuum emission.

\subsection{Blackbody temperature}
\label{sec:tbb}
We begin by exploring the range of best-fitting blackbody temperatures in objects  with redshifts $1.2<z<1.5$.
This redshift range is chosen such that both the {\it WISE W1} and {\it W2} bands are probing the rest frame 1-3\,\micron\ portion of the SED. 
Two independent photometry points in this wavelength range allow us to fit
for both the blackbody temperature and normalisation simultaneously in individual objects.

For this test we keep only objects with $Y<18.1$ which are classed as high confidence point sources in all four UKIDSS bands
(i.e.\,MERGEDCLASS==-1) to avoid objects which are contaminated by host-galaxy emission (see Section~\ref{sec:model} and Appendix~\ref{sec:host}). We require a detection in all of the \textit{grizYJKW12} bands, giving a sample of 1213 quasars.

\begin{figure}
    \includegraphics[width=\columnwidth]{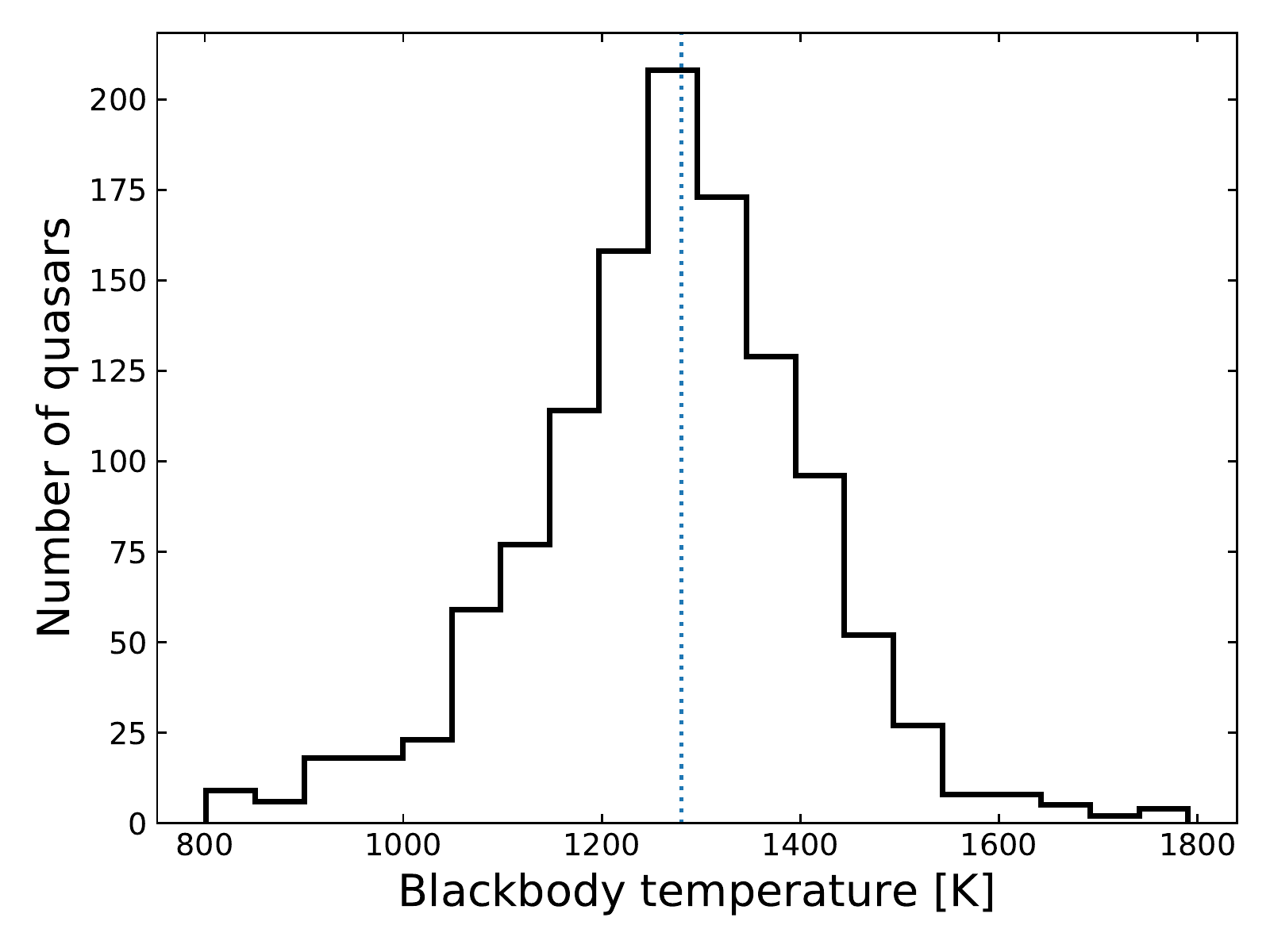}
    \caption{The distribution of blackbody temperatures in our sample of $1.2<z<1.5$, $Y<18.1$ quasars.
    We find that the high-frequency cutoff to the infrared emission (i.e. the 1-3\,\micron\ region) can be well described by a 1280\K\ blackbody.}
    \label{fig:tbb}
\end{figure}

We fit our model to each of these 1213 individual objects with three free parameters:
the $E(B-V)$, the blackbody temperature, and the blackbody strength. The distribution of best fitting blackbody temperatures is shown in Figure~\ref{fig:tbb}. We find that the blackbody temperatures are evenly distributed around the median value of 1280\K.

We then repeat our fitting routine for the same sample, but this time fixing the blackbody temperature to be $T=1280\K$. 
We find that fixing the blackbody temperature reduces the number of objects where the model might be over-fitting the data, while not increasing the number of objects with poor fits: 
for example, the number of quasars for which the $\chisq$ value per degree of freedom lies in the range 0.4-3.0 increases from 1098 to 1123 when fitting this simpler model.

We therefore proceed in the knowledge that the excess rest-frame 1-3\,\micron\ emission in our sample of quasars can be adequately described using a fixed temperature $T=1280\K$ blackbody in our SED model, consistent with emission from dust at its sublimation temperature
(e.g.\,\citealt{Barvainis87}).
However, we ascribe no further significance to the exact value of this temperature, noting instead that it is merely the temperature describing the blackbody which best describes the infrared colours of the quasars, and not necessarily the exact physical temperature of the emitting dust.
In particular, we note that this temperature is  degenerate with the slope of the power law continuum ($-$1.84) which we assume in our model, and a different continuum slope would lead to a slightly different hot dust temperature. Our subsequent results would be qualitatively unchanged.

\subsection{Primary sample}
\label{sec:primary}

Fixing the blackbody temperature allows us to fit our model to higher redshift objects, where we no longer have two independent data points probing the hottest dust emission.

For what we hereafter refer to as our `primary' sample, we take all quasars in the redshift range 
$1.2<z<2.0$ (see Appendix~\ref{sec:highz} for a discussion of the limitations of higher redshift objects).  For our primary sample, the \textit{W2} band probes rest frame wavelengths $\approx$2\,\micron\ where the hot dust blackbody is expected to be the dominant contributor to the total flux. 

We require detections in all of the \textit{JKW12} bands. To avoid biases against hot-dust-faint objects which might drop out of one or more of these bands we restrict our primary sample to $Y<19.3$, where the sample is better than 95\,per cent complete in all UKIDSS bands, giving a total of 12\,294 objects.
We have verified that restricting our sample to a brighter flux limit does not change any of the subsequent results.
When constructing the sample, only 243 objects are lost due to missing unWISE data, meaning our primary sample is more than 98 per cent complete in \textit{W1} and \textit{W2}. Some of these objects will be missed due to crowded fields as the {\it WISE} point spread function is significantly broader than that of SDSS. However, we cannot rule out the possibility that a small fraction of objects are lost due to intrinsically faint fluxes in the rest frame near-infrared. 

We fit our primary sample with two free parameters in the model,
fixing $T_{\rm blackbody}=1280$\K\ and allowing just the $E(B-V)$ and normalisation of the hot dust component to vary.
The median \chisq\ per degree of freedom when fitting our model to this sample is 1.30.
We show in Appendix~\ref{sec:app_tbb} that our subsequent results remain unchanged when instead using a hotter or cooler blackbody in our model quasar SED.
5022 objects from our primary sample with $z>1.56$ also have \civ\ line information from \citet{Rankine20}.
As the sample is restricted to $z<2.0$, \ion{Mg}{II} emission is present in the SDSS spectra to enable black hole mass and Eddington ratio estimation.

\section{Results}
\label{sec:results}

We quantify our results using the ratio of the luminosities at 2\,\micron\  of the hot dust blackbody and extended UV-optical power law continuum in the best fitting model SED for each quasar ($L_{\small 2\,\micron, \textrm{HD}}/ L_{\small 2\,\micron, \textrm{QSO}}$). This ratio is effectively a measurement of the covering factor of the sublimation temperature dust around the UV continuum source. Fitting our model to the primary sample as described above, $L_{\small 2\,\micron, \textrm{HD}}/ L_{\small 2\,\micron, \textrm{QSO}}$ takes values between 0 and $\approx$6, with a median value of 2.5.

We note that $L_{\small 2\,\micron, \textrm{HD}}/ L_{\small 2\,\micron, \textrm{QSO}}=2.5$ corresponds to  log$(L_{2.3\,\micron}/L_{5100\,{\text{\normalfont\AA}}\,})=-0.0064$ for a typical UV continuum slope (i.e., $E(B-V)=0.0$), consistent with the $L_{2.3\,\micron} - L_{5100\,{\text{\normalfont\AA}}}$ relation quoted by \citet{2013ApJ...779..104J} which was derived over a much larger redshift interval than the range considered in this work.
For comparison, the `hot dust poor' criterion of log$(L_{2.3\,\micron}/L_{5100\,{\text{\normalfont\AA}}\,})<-0.5$ used by \citet{2013ApJ...779..104J} is equivalent to $L_{\small 2\,\micron, \textrm{HD}}/ L_{\small 2\,\micron, \textrm{QSO}} \lesssim 0.15$ in our formalism, with a slight dependence on the UV continuum-slope. We find that $\approx$0.1\,per cent of our primary sample satisfy this criterion. However, the number of objects which will scatter to extreme values of the inferred hot dust strength depends strongly on the fractional uncertainties on the photometry (see Appendices~\ref{sec:WISE} and \ref{sec:highz}), and a direct comparison with the `hot dust poor' fraction of  \citet{2013ApJ...779..104J} is not appropriate.

\subsection{Trends with emission line outflow properties}

\begin{figure}
    \includegraphics[width=\columnwidth]{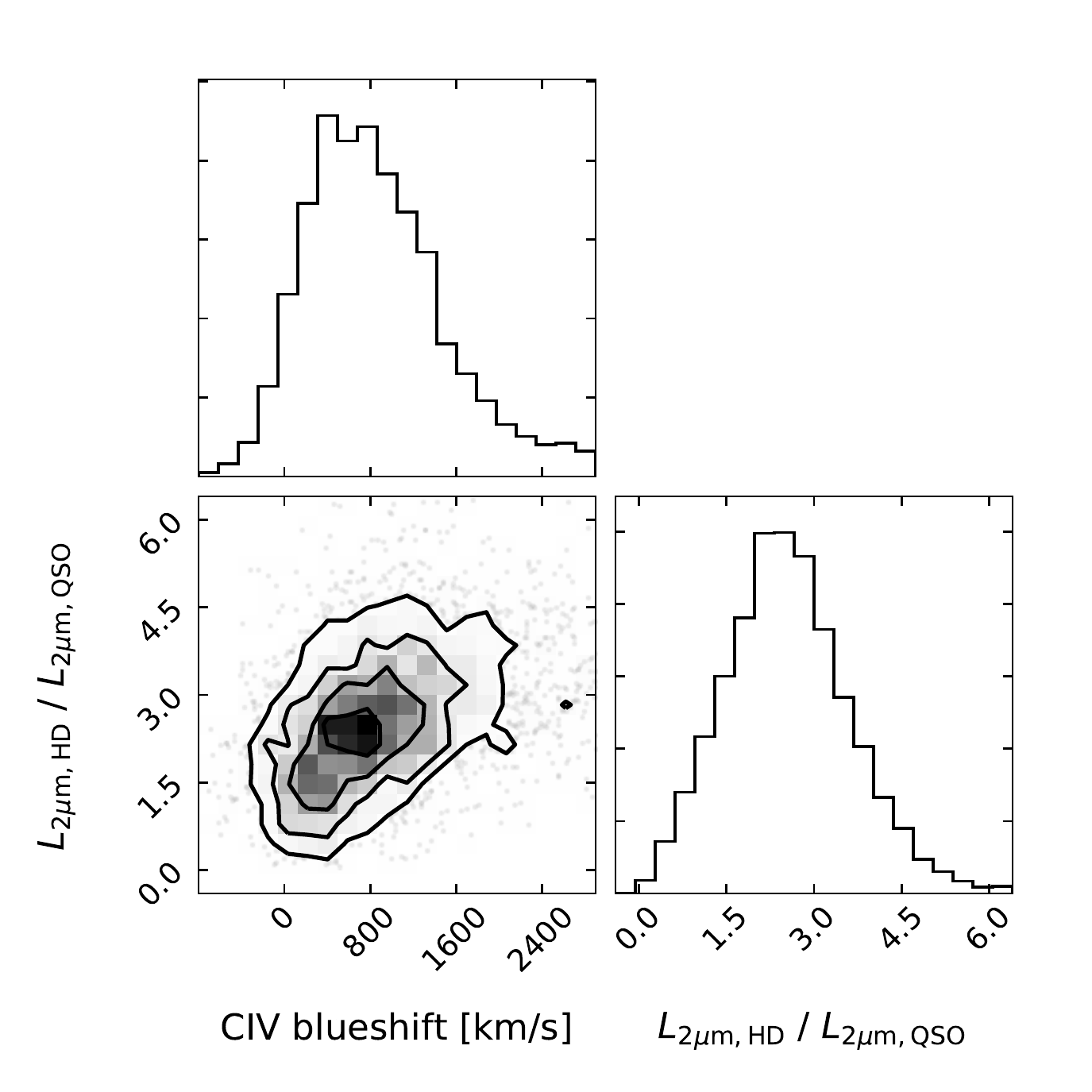}
    \caption{Results of fitting our model to the 5022 quasars from the primary sample with spectral coverage of \civ.
    The bottom left panel shows the distribution of our sample in the plane of \civ\ blueshift and hot dust strength, with points representing individual objects and darker shades representing regions of higher point density. The contours enclose 86, 68, 39 and 12\,per cent of the sample respectively. The normalised 1-D distributions of each parameter are shown in the top left and bottom right.
    Objects with stronger \civ\ outflow signatures tend to have stronger hot dust emission.}
    \label{fig:BBnorm_blue}
\end{figure}

In Fig.~\ref{fig:BBnorm_blue} we show the strength of hot dust emission
as a function of the blueshift of the median flux in the \civ\ emission line relative to the systemic frame. 
There is a clear trend: 
quasars with stronger emission from outflowing ionized gas also show stronger emission from the hottest temperature dust, 
when compared to their UV--optical continuum luminosity,
consistent with the results of \citet{Wang13}.

For objects with \civ\ blueshift less than 200\kmps, the median $L_{\small 2\,\micron, \textrm{HD}}/ L_{\small 2\,\micron, \textrm{QSO}}$ in our primary sample is 1.75, while for those with \civ\ blueshift greater than 2000\kmps, the median $L_{\small 2\,\micron, \textrm{HD}}/ L_{\small 2\,\micron, \textrm{QSO}}$ is 3.09. 
However, objects with fainter observed magnitudes will have larger photometric uncertainties, with the noisier photometry leading to increased uncertainty in our measurement of the hot dust luminosity. We would therefore expect the scatter in Fig.~\ref{fig:BBnorm_blue} to reduce if we restrict our sample to brighter objects.
Taking only the objects with $Y<18.3$, we find the median $L_{\small 2\,\micron, \textrm{HD}}/ L_{\small 2\,\micron, \textrm{QSO}}$ for objects with \civ\ blueshift <200 \kmps\ decreases to 1.50, while the median $L_{\small 2\,\micron, \textrm{HD}}/ L_{\small 2\,\micron, \textrm{QSO}}$ for objects with \civ\ blueshift >2000 \kmps\ increases to 3.14.

The covering fraction of hot dust around the continuum source in our sample of quasars is therefore increasing, on average, by a factor of at least two as we go from objects with \civ\ emission predominantly at the systemic redshift to objects where the kinematics of the \civ\ emission are dominated by outflowing gas.
The presence of a strong nuclear outflow, such as described by a disc-driven wind, must therefore correlate with significant changes in either the geometry or extent (or both) of the hot dust emitting surface.

\subsection{Luminosity and Eddington fraction}

It is known that the \civ\ emission line properties are correlated with other intrinsic quasar properties \citep[][]{Richards11, Rankine20}, with  stronger outflow signatures seen in objects which are generally brighter and have a higher Eddington ratio.
We therefore need to determine whether the variation in the hot dust emission observed in Fig.~\ref{fig:BBnorm_blue} is primarily driven by changes in the outflow properties of the broad line region, or whether it is instead a by-product of a more fundamental correlation.

\begin{figure}
    \includegraphics[width=\columnwidth]{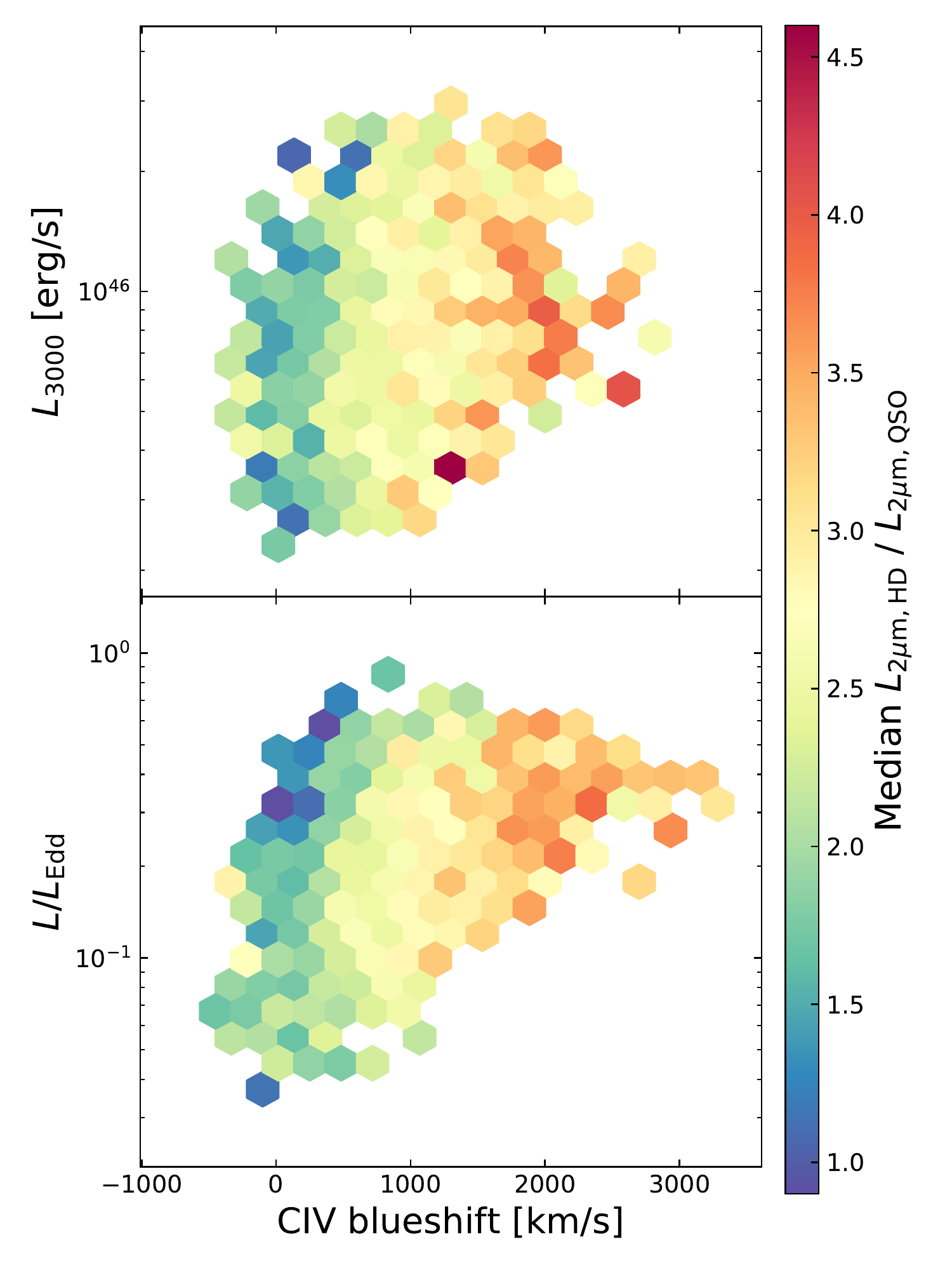}
    \caption{The median hot dust strength in binned regions of parameter space. Only bins with five or more objects are plotted. At fixed values of \civ\ blueshift, the relative strength of hot dust emission is not dependant on either the UV-continuum luminosity or the Eddington ratio, although objects with stronger \civ\ outflow signatures and hence stronger hot dust emission are only found at  $L/L_{\rm Edd}\gtrapprox 0.2$.}
    \label{fig:BBnorm_Ledd}
\end{figure}

In Fig.~\ref{fig:BBnorm_Ledd}, we show the median $L_{\small 2\,\micron, \textrm{HD}}/ L_{\small 2\,\micron, \textrm{QSO}}$ in bins of luminosity, Eddington ratio and \civ\ blueshift. We recover the known relationship between \civ\ blueshift and $L/L_{\rm Edd}$, in that only objects with larger Eddington fractions are observed to have the strongest emission-line outflow signatures. However, when controlling for the observed outflow properties, i.e.\,taking fixed values of \civ\ blueshift, we do not see any trend in the hot dust emission strength as a function of either the Eddington ratio or the UV-continuum luminosity.
There is no correlation between $L_{\small 2\,\micron, \textrm{HD}}/ L_{\small 2\,\micron, \textrm{QSO}}$ and the black hole mass estimated from \mgii.
In other words, the correlation between the strength of the hot dust emission and the blueshift of the \civ\ emission line is not a secondary effect driven by the black hole mass, continuum luminosity, or Eddington ratio.

The relationship between hot dust strength, \civ\ blueshift and Eddington ratio shown in the bottom panel of Fig.~\ref{fig:BBnorm_Ledd} naturally accounts for the varying strength of the correlations detected by \citet{Wang13}: 
if we were to bin our sample by Eddington fraction,
then subsamples with larger $L/L_\textrm{Edd}$ would have an increasing dynamic range in \civ\ blueshift,
and so would display a stronger correlation between the \civ\ blueshift and the strength of hot dust emission.

\subsection{Broad absorption line quasars}
\begin{figure}
    \includegraphics[width=\columnwidth]{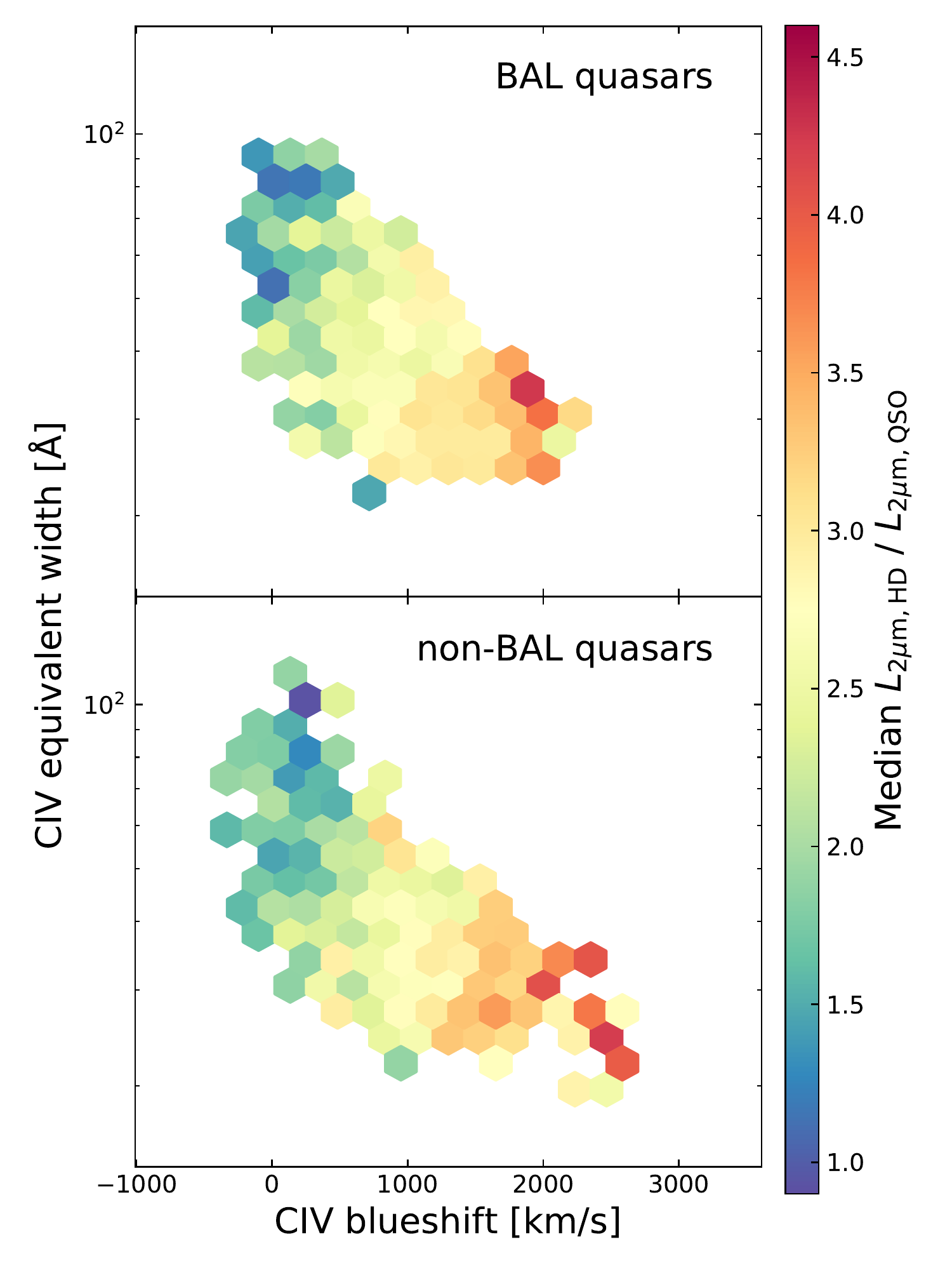}
    \caption{Comparison of the hot dust emission in quasars which show high ionization broad absorption lines (top) and those which don't (bottom). Only bins with five or more objects are plotted. For a given \civ\ blueshift and line equivalent width, significant differences in the median strength of hot dust emission are not seen.}
    \label{fig:BALs}
\end{figure}

Using the spectral reconstructions presented in \citet{Rankine20}, we can  infer unabsorbed \civ\ emission-line properties for those quasars which show high-ionization broad absorption features in their raw spectra (`HiBAL' quasars).
For a given \civ\ blueshift and equivalent width, we find no significant difference in the hot dust emission strengths in HiBAL and non-BAL quasars, as shown in Fig.~\ref{fig:BALs}.

We note that quasars which show larger \civ\ emission-line  blueshifts also show faster and stronger  broad absorption troughs \citep{Rankine20}. We have shown above that the strength of hot-dust emission is correlated with the  \civ\ blueshift, and therefore expect quasars with the most extreme broad-absorption features to also show stronger hot-dust emission, consistent with the results of \citet{2014ApJ...786...42Z}.

\section{Discussion}
\label{sec:discuss}

In the previous section, we showed that the strength of the hot-dust emission in type-1 quasars strongly correlates with the blueshift of the \civ\ emission line. Moreover, when holding the \civ\ blueshift constant, the hot dust emission does not correlate with 3000\,\AA\ luminosity, black hole mass, or the inferred Eddington ratio.

If we assume that the hottest dust emission is coming from material located at the dust sublimation radius, such that it forms a surface (or thin `zone') with only cooler dust components emitting from behind, then the variation in hot dust emission strength could be ascribed to one of two factors.
The total amount of sublimation-temperature dust could be changing, either through an increase in the sublimation radius itself, or through an increase in the dust covering fraction of the sublimation-temperature isotherm.
Alternatively, if the sublimation-temperature dust is forming an optically thick surface, then the change in hot dust strength which we observe could be due to a change in opening angle such that more of the hot dust emitting surface is projected onto the plane of the sky of the observer.

\subsection{`Failed wind' models of the broad  line region}
\label{sec:FRADO}

One theory for the origin of the broad emission line region (BLR) which has gained popularity is that of a `failed wind', where the BLR gas at the systemic redshift is gas  which has failed to reach escape velocity and form a wind, and has instead fallen back to the disc
\citep{2011A&A...525L...8C, 2017ApJ...846..154C, 2018MNRAS.474.1970B}.
The failed wind model is consistent with the first-order behaviour of the broad emission lines, where the equivalent width of lines such as \civ, \ion{C}{III}], Lyman\,$\alpha$, and indeed the Balmer lines, is observed to anticorrelate with the blueshift of the \civ\ line.

However, as discussed by \citet{2019A&A...630A..94G}, a more nuanced version of this picture exists, in which the high- and low- ionization parts of the BLR arise in different locations: the high-ionization lines (such as \civ) arise closer in and form from a failed line-driven wind, whereas the low-ionization lines (such as \mgii) form further out from a failed wind which would otherwise be driven by radiation pressure on dusty grains. This model of failed winds with multiple driving mechanisms is consistent with the picture of the BLR derived from reverberation mapping, which has a stratified velocity structure where high-ionization lines respond to the continuum on shorter time-scales than low-ionization lines.

On first sight, our results would appear to be in slight tension with this model: one natural explanation for the correlation we observe between the hot dust emission and the \civ\ blueshift would be if the hot dust is itself providing the opacity to accelerate the outflow off the disc which we then observe in the ionized gas kinematics.
On the other hand, however the outflow is formed, even if MHD driven \citep{Murray95, 2006ApJ...648L.101E}, we would still expect dust to form inside the wind \citep{2019ApJ...885..126S}. Thus, however the wind(s) are driven, there are perhaps good theoretical reasons to expect that they may correlate with dust emission. 
The fact that we observe this dust to be near the sublimation temperature therefore suggests that, 
if it is not already present and providing the opacity to drive the outflow through radiation pressure,
the dust in the wind must be re-forming at (or near) a height above the disc corresponding to the limit of the  dust sublimation region.

\subsection{UV continuum slope}
\label{sec:UV}

\begin{figure*}
    \includegraphics[width=2\columnwidth]{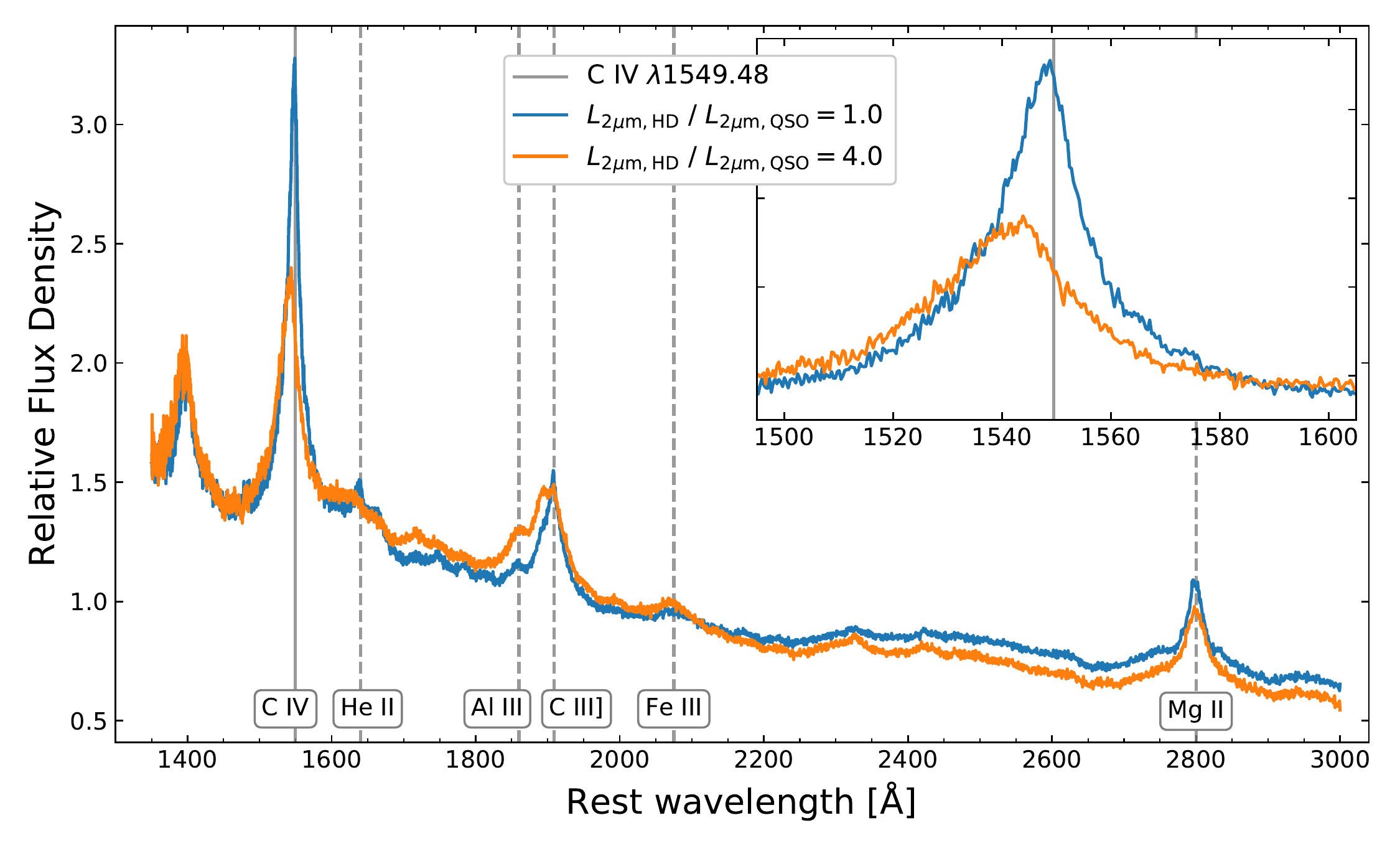}
    \caption{Composite spectra for objects with extremely strong (weak) hot dust emission, plotted in orange (blue). 
    {Each composite contains 84 objects, and the mean accretion rate in both composites is $L/L_\textrm{Edd}=0.26$.}
    The rest-frame wavelength of the \civ\ emission line is marked in solid gray. Quasars with stronger hot dust emission tend to have stronger emission in the blue wing of the \civ\ line and weaker emission at the systemic wavelength, combining to give an asymmetric and blueshifted line profile indicative of outflowing gas. Such objects tend to have a bluer UV continuum, suggesting that the hot dust is not obscuring the central source. The quasars with weaker hot dust emission have stronger emission from other broad lines such as \ion{He}{II} $\lambda$1640 and \ion{C}{III}] $\lambda$1909, consistent with a scenario in which part of the broad emission line region is produced from gas which fails to form an outflowing wind.
    }
    \label{fig:comp_spectra}
\end{figure*}

In Fig.~\ref{fig:comp_spectra}, we show stacked spectra in bins of hot dust strength. 
The two bins have been chosen to be matched in Eddington fraction: each quasar contributing to the composites has $-1.0<\textrm{log}(L/L_\textrm{Edd})<-0.2$ and the average accretion rate in each composite is $L/L_\textrm{Edd}=0.26$. Each bin contains 84 objects.

We can see that quasars with stronger hot dust emission tend to have slightly bluer UV continua on average, consistent with their observed photometry: quasars with $L_{\small 2\,\micron, \textrm{HD}}/ L_{\small 2\,\micron, \textrm{QSO}}\simeq 1$ have a median  colour $g-K=2.32$, while quasars with  $L_{\small 2\,\micron, \textrm{HD}}/ L_{\small 2\,\micron, \textrm{QSO}}\simeq 4$ have a median colour $g-K=1.98$. Equivalently, our best fitting SED models (Section~\ref{sec:model}) have average $E(B-V)$s of $0.020$ and $-0.017$ respectively. We would expect any dust obscuration to redden the continuum, and so the fact that quasars with stronger hot dust emission are observed to have slightly bluer UV--optical continua suggests that the emitting hot dust does not intersect the line of sight to the accretion disc.
This finding is consistent with the results of \citet{Kim18}, who find that the hot dust emission is not significantly stronger in dust-reddened quasars (with $E(B-V)\approx1$) when compared to their unobscured counterparts.

The correlation between SED slope and hot dust strength is consistent with known trends with the \civ\ line, where objects with larger emission line blueshifts tend to have bluer UV continua \citep{Richards11, Rankine20}.
We can also see from Fig.~\ref{fig:comp_spectra} that the objects with stronger hot dust emission also have stronger \ion{Al}{III}\,$\lambda$1860 and \ion{Fe}{III}\,$\lambda$2075,
consistent with the results of \citet{2020MNRAS.496.2565T} which showed that quasars with stronger \civ\ outflow signatures show stronger emission from dense gas close-in to the central ionizing source.

\subsection{Bolometric corrections and EUV flux}
\label{sec:BC}

Estimating an Eddington fraction requires knowledge of the bolometric luminosity of the quasar, and we have shown that some quasars show significantly more emission in the near infrared than others of the same 3000\,\AA\ luminosity. 
It is true that objects with a larger $L_{\small 2\,\micron, \textrm{HD}}/ L_{\small 2\,\micron, \textrm{QSO}}$ for a given 3000\,\AA\ luminosity will have a larger contribution to their bolometric luminosity (and thus their Eddington fraction) from the rest-frame near infrared.
However, a larger uncertainty in the inference of a bolometric luminosity arises from the `unseen' extreme ultraviolet (EUV) continuum
\citep[e.g.,][]{Krawczyk13}. 
There are also inherent uncertainties in the estimation of black hole masses of order 0.5\,dex, and so the uncertainty in $L/L_{\rm Edd}$ in Fig.~\ref{fig:BBnorm_Ledd} arising from the variation in the strength of the hot dust emission  is sub-dominant compared to other uncertainties in that measurement.

Indeed, it has been suggested that the ratio of EUV flux to optical flux varies as a function of Eddington fraction,  with higher accretion rate AGN having stronger EUV emission by factors of at least a few \citep[e.g.][]{2012MNRAS.425..907J}.
If this change in EUV flux is responsible for driving the outflows inferred from the \civ\ line profiles, then it is possible that the relative strength of the EUV flux might be responsible for the correlation between \civ\ blueshift and hot dust emission reported in this work.
Increasing the amount of EUV flux (at fixed $L_{3000}$) would increase the amount of energy which is available to heat the dust and  therefore increase the strength of hot dust emission which we measure relative to the optical continuum.
The relative weakness of \ion{He}{II} emission seen in Fig.~\ref{fig:comp_spectra},
despite the hypothesised increase in EUV flux, could then be explained via a decreasing covering fraction of line-emitting clouds with increasing Eddington fraction, as discussed by \citet{2020MNRAS.494.5917F}.
In this scenario, the assumption of a constant bolometric correction which we have made throughout this work would clearly be wrong, and the Eddington fractions which we have inferred would be biased high in low-blueshift objects and biased low in high-blueshift objects.

\subsection{Comparison with Jiang et al. (2010)}

\citet{Jiang10} reported the discovery of two quasars (from a sample of 21) at redshift $z>6$ with very weak hot dust emission, and suggested that these were `first-generation quasars born in dust-free environments' in the  early universe, i.e. quasars which were too young to have been able to form dust around them, or at least enough dust to be detected in the rest-frame near-infrared.
We note that these two quasars,  J0005-0006 and J0303-0019, both show very narrow, symmetric \civ\ emission line profiles, as can be seen from their spectra \citep{2007ApJ...669...32K, 2009ApJ...702..833K}.
In our parametrization, these two objects would lie in the bottom-left corner of Fig.~\ref{fig:BBnorm_blue}, and we believe that these two quasars are consistent with the trend between the hot dust emission and the emission line properties which we have identified at $z\le2$. 
Our results would therefore imply that the unusual infrared properties of these $z>6$ objects are driven by the shape of their ionizing SEDs, or whatever else is responsible for their (lack of) BLR outflows, and not by the dust content of their environments.

\subsection{Hot dust reverberation}

Reverberation mapping (RM) experiments have found that the time lag in the response of the hot dust emission to variations in the optical continuum ($\tau$) is  well enough correlated with the AGN luminosity to lead to the idea of using the hot dust reverberation in AGN as a standardisable candle which could be used to test and constrain cosmological models 
\citep[e.g.][]{2014ApJ...784L...4H, 2017MNRAS.464.1693H}.

The results of this paper show that there are systematic changes in either the geometry or extent (or both) of the sublimation-temperature dust in luminous quasars as a function of their ionized gas outflow properties.
If the variable component of the hot dust emission is the same as the `static' emission quantified in this work,
then these changes could potentially introduce a \civ-blueshift dependant bias into the lag-luminosity relation which is used to estimate the intrinsic brightness of the AGN.
However, theoretical work has shown  that the infrared response function is not very sensitive to the exact distribution of the reverberating dust  \citep{2020ApJ...891...26A}.
RM experiments to date have found a small scatter in the lag-luminosity relation, which does not appear to increase with Eddington ratio \citep[][fig. 10]{2020ApJ...900...58Y},
suggesting that any outflow-dependant changes in the hot dust reverberation properties are small compared to other sources of scatter in the $\tau-L$ relation. 
While there is therefore no direct evidence that the $\tau-L$ relation is affected by any outflow properties, we note that the majority of hot dust RM experiments have so far been conducted on local  Seyfert galaxies with $L/L_{\rm Edd}<0.3$ \citep{2006ApJ...639...46S, 2014ApJ...788..159K}, where we do not expect to see large \civ\ blueshifts.

\subsection{Far-infrared properties}

Debate has raged over the origin of the far-infrared (FIR) emission in luminous AGN.
As the emission from the toriodal obscuring structure is insufficient to account for the observed FIR flux, it is usually argued that the FIR emission is predominantly due to dust heated by young stars and supernovae, and thus used as an indicator of the star formation rate 
\citep[e.g.][and references therein]{2016MNRAS.457.4179H}.
However, such arguments are based on the assumption that all quasars have the same SED, which is not true \citep[e.g.][]{Krawczyk13}. 
The  question then arises as to whether or not the diversity of AGN SEDs could encompass variation in the longer wavelength FIR emission, especially  in objects with evidence for AGN-driven winds. 
For example, \citet{Symeonidis17}  argued that the FIR emission in $z\approx2$ quasars is instead due to dust on kilo-parsec scales which is being heated by the AGN.  

We note that any AGN-driven winds are likely to correlated across many scales. For example, \citet{Coatman19} found that the \civ\ blueshift is correlated with the outflow signature in the [\ion{O}{III}] $\lambda$5008 emission line, which is believed to trace ionized gas on larger, possibly galaxy-wide, scales.
To simplify the following qualitative discussion, we therefore  do not distinguish between AGN-driven outflows  on different scales, but instead consider a toy model where objects either do or do not show evidence for what we hereafter refer to as `winds'.

Within this model, quasars at $z\approx2$ which are actively driving winds are expected to show larger \civ\ blueshifts in their spectra. 
\citet{Maddox17} showed that the individual FIR detection rate of such quasars is higher than would be expected for their optical luminosities.
More specifically, objects which are detected in the \textit{Herschel}-ATLAS survey at 100-500\,\micron\ are disproportionately more likely to show large \civ\ blueshifts. 
Equivalently, quasars with large \civ\ blueshifts have weaker \civ\ equivalent widths, and \citet{2016MNRAS.457.4179H} showed that these objects have stronger \textit{Herschel} emission when stacking optically selected objects in order to quantify the FIR emission below the individual-object detection limit.

More recently, \citet{Baron19} have shown that type 2 AGN at $z\approx0.1$ with outflow signatures in [\ion{O}{III}] show an excess of emission in the mid-infrared ($3\,\micron<\lambda<20\,\micron$). 
Together with the results in this work, that objects with stronger winds have brighter near-infrared emission, a picture is therefore emerging where  objects with stronger winds show stronger emission from dust across a wide range of temperatures. 
However, while previous works have attributed the brighter FIR emission in objects with stronger winds  to increased rates of star formation, the stronger emission from hot, near-sublimation temperature dust which we report in this work cannot be ascribed to dust which is being heated by young stars.  The evidence  we have  presented that quasar-driven outflows are affecting the emission from AGN-heated dust structures therefore provides support for the idea that the increased FIR emission in windy quasars has a significant component due to AGN-heated dust and is not solely due to increased rates of star formation.

This scenario is also consistent with the results of \citet{2014ApJ...785..154L}
who showed that FIR-bright quasars have weaker H\,$\alpha$ and Ly\,$\alpha$ emission, and that objects with FIR detections are preferentially brighter in the rest-frame near-infrared. In the models discussed in Section~\ref{sec:FRADO}, quasars which fail to drive strong winds have larger amounts of gas falling back to form the BLR and hence emit stronger broad lines. These objects are also less likely to drive galaxy-scale outflows which in turn removes a source of AGN-heated dust and they thus possess fainter FIR emission.

We note in passing that, whatever is driving the variation in the FIR emission, the fact that the FIR emission correlates with the \civ\ blueshift means that  FIR-selected samples are potentially biased in ways which are not well understood. 
In particular, 
we have shown that the hot dust strength is also correlated with the \civ\ blueshift, and thus care must be taken when reporting hot dust properties of samples which are selected using their FIR emission properties. 
For example, the `Hot Dust Poor' fraction, i.e. the percentage of the population which are underluminous in the near-infrared \citep[e.g.][]{Hao10, Hao11}, could be biased in samples which are selected using their FIR properties.

\section{Summary}

Using photometric data from SDSS, UKIDSS-LAS and unWISE, we have investigated the properties of the hot dust emission in luminous type 1 quasars across the redshift range $1.2<z<2.0$.
We find that the 1-3\,\micron\
excess in quasars can be well described by a  single temperature blackbody, consistent with emission from sublimation temperature dust. We quantify the strength of the hot dust emission as the normalisation of this blackbody relative to the UV--optical power-law continuum at 2\,\micron. Our main results are:

\begin{itemize}
    \item The strength of the hot dust emission correlates with the strength of the blueshift of the  \civ\ emission line, in agreement with \citet{Wang13}.
    In a disc-wind model, this correlation suggests that objects with more wind-dominated emission have a larger surface of sublimation-temperature dust visible to the observer.
    
    \item When controlling for the emission line properties, and assuming a constant bolometric correction, the strength of the hot dust emission does not vary with UV luminosity, black hole mass, or Eddington fraction.
    In the picture presented by \citet{2019A&A...630A..94G}, this would imply that the main driver of the wind, whether increased $M_{\rm BH}$ at lower $L/L_{\rm Edd}$, or increased $L/L_{\rm Edd}$ at lower $M_{\rm BH}$, does not affect the emission from hot dust.
   
    \item Stacking the rest-frame UV spectra of these quasars, we find there is a small but significant change in the average slope of the UV continuum with varying hot dust emission strength: objects with stronger hot dust emission tend to have bluer continua.
    The parity of this change in slope suggests that the sublimation-temperature dust is not located along the line of sight towards the source of the UV continuum, and is consistent with previously reported correlations between the quasar SED and the quasar emission-line properties.
    
    \item There appears to be no difference in the hot dust properties of BAL and non-BAL quasars at given \civ\ line properties, consistent with the result found by \citet{Rankine20} that BALs and non-BALs are drawn from the same parent population.
    
\end{itemize}

\noindent 
Outflows off the accretion disc could be driven by radiation pressure on dusty gas, or by other mechanisms, and our results are not able to rule out any specific mechanism. 
However the winds are formed, they do correlate with the properties of the hottest emitting dust. Dust of this temperature cannot be heated by star formation, and our results therefore provide evidence of a physical link between the properties of the broad emission line region and properties of the dust emitting regions. 
Moreover, this link calls into question the common assumption that the far-infrared emission in quasars is due to star formation alone and not due to dust which has been heated by the AGN.

\section*{Acknowledgements}


We thank  Triana Almeyda, Matt Auger, Chris Done, Gary Ferland, Seb H\"onig, Christian Knigge, Knox Long, James Matthews, Nadia Zakamska and an anonymous referee for useful comments.
MJT and ALR thank the Science and Technology Facilities Council (STFC) for the award of studentships. MJT is grateful to the Royal Astronomical Society, the Cambridge Philosophical Society and St Catharine's College for the award of travel grants.
MB acknowledges funding from the Royal Society via a University Research Fellowship.
PCH acknowledges funding from STFC via the Institute of Astronomy, Cambridge, Consolidated Grant.

This work  made use of AstroPy \citep{astropy:2013, astropy:2018}, corner.py \citep{corner}, Matplotlib \citep{Hunter:2007}, NumPy \citep{numpy}, SciPy \citep{scipy} and Q3C \citep{Q3C}.
This work also made use of the Whole Sky Database (wsdb) created by Sergey Koposov and maintained at the Institute of Astronomy, Cambridge by Sergey Koposov, Vasily Belokurov and Wyn Evans with financial support from STFC and the European Research Council. 
We thank M. Vestergaard for making the  Iron template available.
This research has made use of the SVO Filter Profile Service (http://svo2.cab.inta-csic.es/theory/fps/) supported from the Spanish MINECO through grant AYA2017-84089.

This publication makes use of data products from the Wide-field Infrared Survey Explorer, which is a joint project of the University of California, Los Angeles, and the Jet Propulsion Laboratory/California Institute of Technology, and NEOWISE, which is a project of the Jet Propulsion Laboratory/California Institute of Technology. WISE and NEOWISE are funded by the National Aeronautics and Space Administration.

Funding for the Sloan Digital Sky Survey IV has been provided by the Alfred P. Sloan Foundation, the U.S. Department of Energy Office of Science, and the Participating Institutions. SDSS-IV acknowledges
support and resources from the Center for High-Performance Computing at
the University of Utah. The SDSS web site is www.sdss.org.

SDSS-IV is managed by the Astrophysical Research Consortium for the 
Participating Institutions of the SDSS Collaboration including the 
Brazilian Participation Group, the Carnegie Institution for Science, 
Carnegie Mellon University, the Chilean Participation Group, the French Participation Group, Harvard-Smithsonian Center for Astrophysics, 
Instituto de Astrof\'isica de Canarias, The Johns Hopkins University, Kavli Institute for the Physics and Mathematics of the Universe (IPMU) / 
University of Tokyo, the Korean Participation Group, Lawrence Berkeley National Laboratory, 
Leibniz Institut f\"ur Astrophysik Potsdam (AIP),  
Max-Planck-Institut f\"ur Astronomie (MPIA Heidelberg), 
Max-Planck-Institut f\"ur Astrophysik (MPA Garching), 
Max-Planck-Institut f\"ur Extraterrestrische Physik (MPE), 
National Astronomical Observatories of China, New Mexico State University, 
New York University, University of Notre Dame, 
Observat\'ario Nacional / MCTI, The Ohio State University, 
Pennsylvania State University, Shanghai Astronomical Observatory, 
United Kingdom Participation Group,
Universidad Nacional Aut\'onoma de M\'exico, University of Arizona, 
University of Colorado Boulder, University of Oxford, University of Portsmouth, 
University of Utah, University of Virginia, University of Washington, University of Wisconsin, 
Vanderbilt University, and Yale University.

\section*{Data availability}

The data underlying this article were accessed from the Sloan Digital Sky Survey,\footnote{https://www.sdss.org/dr14/}
the UKIRT Infrared Deep Sky Survey,\footnote{http://wsa.roe.ac.uk}
AllWISE\footnote{https://irsa.ipac.caltech.edu/Missions/wise.htm}
and unWISE\footnote{http://catalog.unwise.me}.
The derived data generated in this research will be shared on reasonable request to the corresponding author.




\bibliographystyle{mnras}
\bibliography{Paper_refs} 




\appendix

\section{SDSS selection effects}
\label{sec:select}

The `core' eBOSS quasar sample \citep{Myers15} makes up a significant fraction of the SDSS-IV quasar population, and thus the SDSS DR14 quasar catalogue. This sample was in part selected using an optical--infrared colour cut, meaning that `core' eBOSS quasars have a force-photometered weighted-average flux in \textit{WISE} \textit{W1} and \textit{W2} above some threshold relative to the optical fluxes from the Sloan imaging data, which potentially biases our sample against quasars with weaker hot dust emission.

To test the effect of this selection criterion on our results, we exclude objects with the EBOSS\_TARGET1 selection flag solely set to QSO1\_EBOSS\_CORE (i.e., EBOSS\_TARGET1 == 2**10). This cut reduces our sample to the SDSS DR9 \citep{Ross12}, which predates \textit{WISE}, while still keeping all objects from SDSS-IV which were targeted through other programs. The total size of the DR14 quasar catalogue is reduced from 526\,356 to 442\,231 objects using this criterion.

Restricting EBOSS\_TARGET1 != 2**10 cuts our primary sample from 5022 to 4635 objects.
The results and conclusions of the paper remain unchanged.

\section{Comparing unWISE and AllWISE}
\label{sec:WISE}

\begin{figure}
\begin{center}
\includegraphics[trim=0 68 10 160,clip,width=\columnwidth,keepaspectratio]{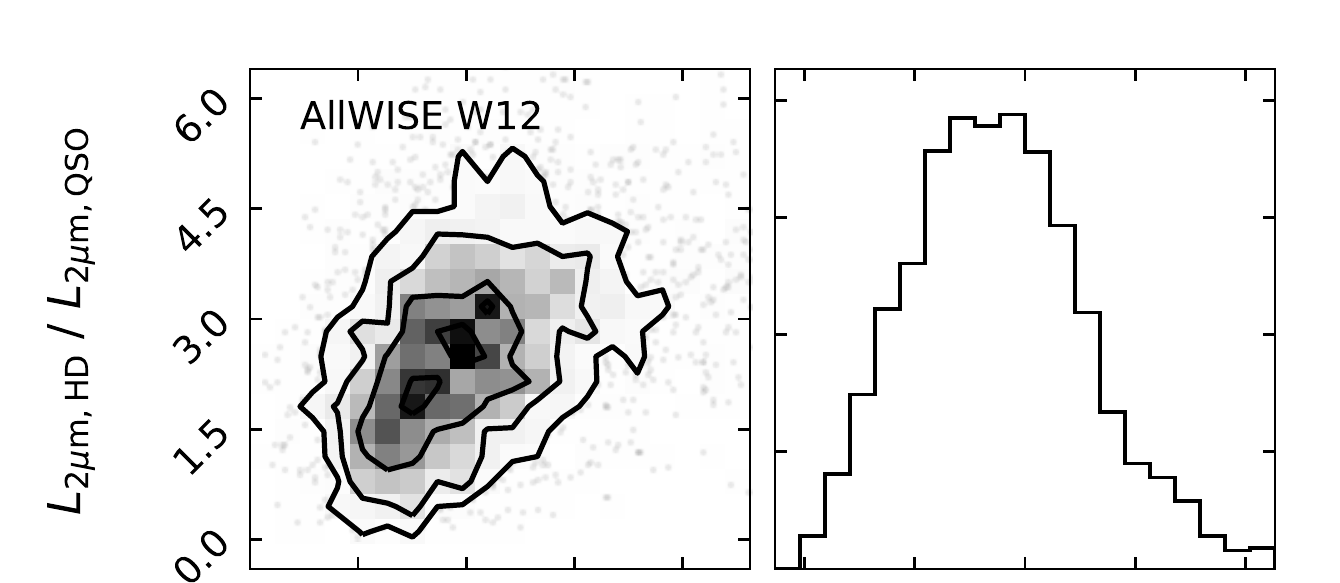}
\includegraphics[trim=0 10 10 175,clip,width=\columnwidth,keepaspectratio]{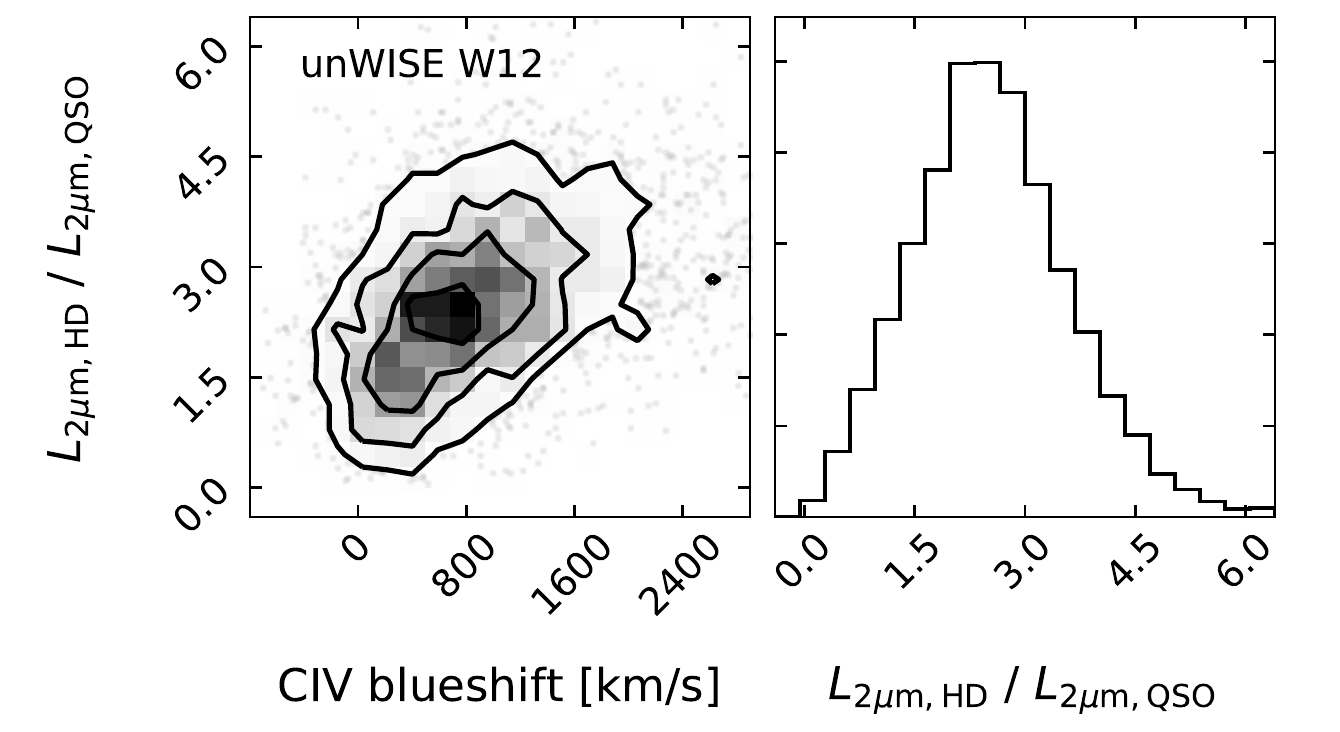}
\end{center}
\caption{Comparison of the strength of hot dust in the primary sample when using AllWISE \textit{W1} and \textit{W2} (top) instead of unWISE data (bottom). On the left is the distribution of $L_{\small 2\,\micron, \textrm{HD}}/ L_{\small 2\,\micron, \textrm{QSO}}$ as a function of \civ\ blueshift, and on the right is the normalised 1-D distribution. Using the deeper photometric data from unWISE leads to a reduction in the scatter in best-fitting model parameters.}
\label{fig:WISE}
\end{figure}

In Fig.~\ref{fig:WISE}, we compare the effect of using 
AllWISE {\it W1} and {\it W2} photometry in place of unWISE when fitting to our primary sample. 

For AllWISE, we use only data points where the signal-to-noise ratio (S/N) is greater than 2 and  the contamination and confusion flags are set to zero in all bands, indicating that the source is not affected by any known image artifacts. As with unWISE, the AllWISE catalogue is matched to SDSS using a 3.0\,arcsec matching radius and we keep only sources with unique matches within that radius. 
The number of DR14 quasars we match to AllWISE  is 397\,873, slightly less than given by \citet{Paris18}
because of the more stringent quality cuts we have imposed on the data.

The unWISE catalogue presented by \citet{Schlafly19}
has $\approx$0.7 magnitudes deeper coverage in {\it W1} and {\it W2} compared to AllWISE, corresponding to additional data from the reactivation of NEOWISE. 
The vast majority of our primary sample are detected in both catalogues, but the additional depth of unWISE corresponds to reduced uncertainty in the {\it W1} and {\it W2} photometry, which leads to a noticeable reduction in the scatter of the normalisation of the best-fitting hot dust blackbody.

While all the results hold true whether AllWISE or unWISE is used, we have chosen to use unWISE data where possible to improve the detectability of trends in our results. AllWISE has only been used for \textit{W3} data when fitting to our `higher redshift' sample in Appendix~\ref{sec:highz}.

\section{Host galaxy flux}
\label{sec:host}

The emission from the  rest-frame 1-3\,\micron\ portion of the galaxy SED is largely dominated by flux from older, cooler stellar populations, and hence it is extremely unlikely the emission is correlated with the AGN outflow properties when the quasar duty cycle is much shorter ($\sim$$10^6$ years) than the age of these stars. Moreover, the flux $\lambda F_\lambda$ from starlight is reasonably flat across the $\simeq0.7$-2\,\micron\ portion of the SED, irrespective of the galaxy type \citep[e.g.][]{SWIRE}, and it is very hard to explain the variation in the strength of the 1-3\,\micron\ excess observed in luminous quasars as due to any variation in the host galaxy.
In other words, the changes in the slopes of the optical and near-infrared parts of the total SED are very different when varying the strength of the hot dust emission compared to those arising from a variable fractional host galaxy contribution, as shown by the `mixing diagrams' of \citet{2013MNRAS.434.3104H}.
We are therefore confident in interpreting the variation in the 1-3\,\micron\ emission as arising from the sublimation-temperature dust around the AGN.

However, while we take steps to mitigate against the contamination of the rest-frame 0.7-3\,\micron\ region by starlight from the host galaxy (i.e. requiring point-source detections in all UKIDSS bands), the possibility remains that the strength of hot dust emission is affected by some relatively small amount of flux from the host galaxy.
We have therefore re-run our fitting routine, including an Sb galaxy template from the SWIRE library \citep{SWIRE} in our SED model with a fixed normalisation such that the integrated galaxy flux in the wavelength interval 4000-5000\,\AA\ is 10 per cent of the integrated quasar flux across the same interval. For a model with an average amount of hot dust emission, this normalisation corresponds to a flux at 1\,\micron\ from starlight  0.5 times that of the flux due to the quasar.

The exact shape of the host galaxy SED does not matter; the desired effect is to add flux to the 1\,\micron\ inflexion-point of the total quasar SED. For the range of quasar extinctions in our sample of SDSS objects (i.e. $E(B-V) \lesssim 0.2$), our prescription is such that the starlight component has a negligible effect on the bluer parts of the total SED. 

Including this additional galaxy-component in our model, we find that the median blackbody temperature (Section~\ref{sec:tbb}) decreases from 1280 to 1190\kelvin, and adopt this temperature as a fixed parameter in the same way as described in the main text.
The inclusion of the galaxy template has the effect of flattening the 1\,\micron\ inflexion point in the total model SED for any given strength of hot dust emission.
Combined with the difference in shape of the slightly cooler blackbody, this results in an increase in the normalisation at 2\,\micron\ of the  best-fitting hot dust by a constant factor of about 50\,per cent. 
The results and conclusions of the paper remain unchanged.


\section{Redshift evolution}
\label{sec:highz}

To investigate the hot dust properties in the quasar population at higher redshifts, it is necessary to make use of data at longer observed wavelengths. 
We have experimented with using the AllWISE \textit{W3} band to constrain the properties of the hot dust emission.
However, \textit{W3} is a much broader band than either \textit{W1} or \textit{W2} (see Fig.~\ref{fig:model}) limiting us to a narrow redshift range where \textit{W3} is not covering any wavelengths longer than 3\,\micron\ in the rest frame, while also not covering any wavelengths shorter than 1\,\micron.

We take all objects with $2.75<z<3.25$ and \textit{rizYJHKW123} data.
We exclude \textit{g} due to Lyman suppression in the IGM at these redshifts.
AllWISE \textit{W3} is not as deep as the shorter  wavelength bands and a brighter flux limit of $Y<18.3$ is necessary to ensure the sample is more than 95\,per cent complete in \textit{W3}. 
The selection gives a total of 701 objects, where we use unWISE as the source of \textit{W1} and \textit{W2} data.
Allowing the blackbody temperature to vary in the same way as described in Section~\ref{sec:tbb} also gives a median $T_{\rm blackbody}$ of 1280\K\ for this higher-redshift sample, and we fit our model using the same parameters as for the primary sample.

We find that the median strength of hot dust emission  $L_{\small 2\,\micron, \textrm{HD}}/ L_{\small 2\,\micron, \textrm{QSO}}$ in this higher-redshift sample is 2.53, very similar to that of the primary sample. The strength of the hot dust emission in our $z\simeq3$ sample also shows a significant correlation with the blueshift of the \civ\ emission line.
However, the difference in filter response curves between \textit{W2} and \textit{W3}, coupled with the different noise properties of each photometric band, combine to give a significant difference in the scatter around the intrinsic hot dust strength. 
For example, a one-sigma change in the \textit{W2} measurement for a $z=2$ object will have a different effect on the best-fitting hot dust strength than that resulting from a one-sigma change in the \textit{W3} measurement for a $z=3$ object.
For this reason, while the average amount of hot dust in our $z\simeq3$ sample appears to be the same as in the primary sample, a detailed comparison of the distribution of hot dust properties within each redshift bin is not appropriate.

\section{Blackbody temperature}
\label{sec:app_tbb}

\begin{figure}
\begin{center}
\includegraphics[trim=0 68 10 160,clip,width=\columnwidth,keepaspectratio]{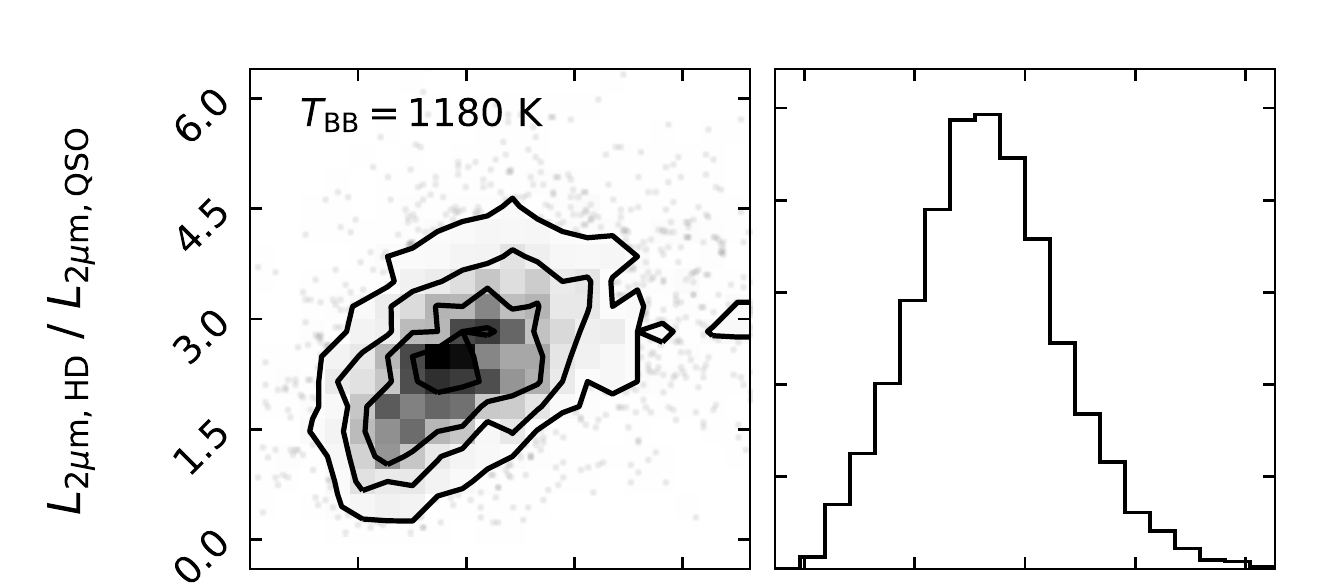}
\includegraphics[trim=0 10 10 175,clip,width=\columnwidth,keepaspectratio]{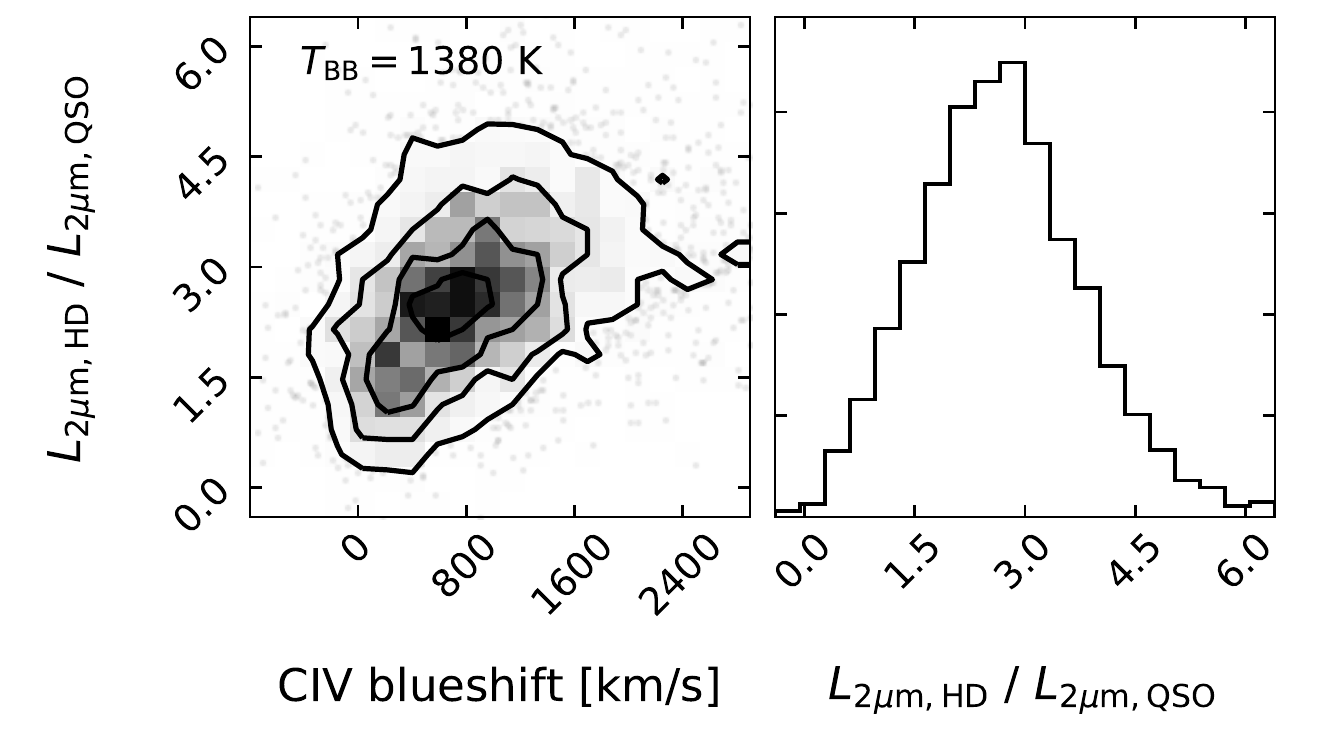}
\end{center}
\caption{Comparison of the strengths of hot dust in the best fitting models when using a blackbody temperature of 1180\kelvin\ (top) or 1380\kelvin\ (bottom).
On the left is the distribution of $L_{\small 2\,\micron, \textrm{HD}}/ L_{\small 2\,\micron, \textrm{QSO}}$ as a function of \civ\ blueshift, and on the right is the normalised 1-D distribution.
Our results  (c.f. our adopted 1280\kelvin\ model in Fig.~\ref{fig:BBnorm_blue}) are not sensitive to the exact form of the blackbody we use to model the SED.}
\label{fig:vary_tbb}
\end{figure}

The effect of changing the temperature of the blackbody used to model the hot dust emission is shown in Fig.~\ref{fig:vary_tbb}.
Using 1180\kelvin\ or 1380\kelvin\
instead of our adopted 1280\kelvin\  as the temperature of the blackbody in our SED model results in an increase in the median \chisq\ per degree of freedom  from 1.30 to 1.33 in each case, but our results are not sensitive to the exact temperature adopted.
In short, if we were to adopt a different temperature blackbody to describe the near-infrared rest frame emission in our model SED, 
the conclusions of the paper would remain unchanged.

\bsp    
\label{lastpage}
\end{document}